\documentclass[letterpaper, 10 pt, conference]{ieeeconf}  

\IEEEoverridecommandlockouts                              

\usepackage{graphics} 
\usepackage{amsmath} 
\usepackage{amssymb}  
\usepackage{enumerate}
\usepackage{mathtools}
\usepackage{graphicx}
\usepackage{graphicx,lipsum}
\usepackage[noadjust]{cite}

\newtheorem{thm}{\textbf{Theorem}}
\newtheorem{lem}{\textbf{Lemma}}

\newtheorem{cor}{\textbf{Corollary}}
\newtheorem{defn}{\textbf{Definition}}

\newtheorem{rem}{\textbf{Remark}}

\title{\LARGE \bf
Multiple Control Barrier Functions: An Application to Reactive Obstacle Avoidance for a Multi-steering Tractor-trailer System
}

\author{Mohammad Aali and Jun Liu
\thanks{This work was supported in part by Nutrien Ltd. and the NSERC of Canada.}
\thanks{Mohammad Aali is with the Department of Applied Mathematics, University of Waterloo, Canada,
        {\tt\small mohammad.aali@uwaterloo.ca}}%
\thanks{Jun Liu is with the Department of Applied Mathematics, University of Waterloo, Canada,
        {\tt\small j.liu@uwaterloo.ca}}%
}

\begin{document}

\maketitle
\thispagestyle{empty}
\pagestyle{empty}

\begin{abstract}
Control barrier functions (CBFs) recently introduced a systematic way to guarantee the system's safety through set invariance. Together with a nominal control method, it establishes a safety-critical control mechanism. The resulting safety constraints can be enforced as hard constraints in quadratic programming (QP) optimization, which rectifies the nominal control law based on the set of safe inputs. In this work, we introduce a multiple CBFs scheme which enforces several safety constraints with high relative degrees. This control structure is essential in many challenging robotic applications that need to meet several safety criteria simultaneously. In order to illustrate the capabilities of the proposed method, we have addressed the problem of reactive obstacle avoidance for a class of tractor-trailer systems. Safety is one of the fundamental issues in autonomous tractor-trailer systems design. The lack of fast response due to poor maneuverability makes reactive obstacle avoidance difficult for these systems. We develop a control structure based on a multiple CBFs scheme for a multi-steering tractor-trailer system to ensure a collision-free maneuver for both the tractor and trailer in the presence of several obstacles. Model predictive control is selected as the nominal tracking controller, and the proposed control strategy is tested in several challenging scenarios.
\end{abstract}

\section{INTRODUCTION}\label{sec1}
CBFs have recently shown a great potential to ensure the safety of dynamical systems in the context of set invariance. In this method, the notion of safety is defined as the ability of the control system to generate a control sequence that avoids entering an unsafe region. Initially, CBFs were introduced for systems with relative degree one \cite{ames2014control}. The QP optimization scheme introduced in \cite{romdlony2014uniting} makes CBFs suitable to incorporate with performance-driven control methods. However, an extension to the systems with higher relative degrees \cite{nguyen2016exponential} broadens its applications from safety-critical control of teams of quadrotors \cite{wang2017safe} to complex dynamical systems such as legged robots \cite{grandia2021multi}. Stochastic, adaptive, and robust safety-critical control methods with CBF have been developed recently \cite{Wang2021, Xiao2021, Liu2021}.

The structure of real-world safety-critical systems (e.g. multi-body mobile robots) implies that a single CBF is not enough for guaranteeing the safety of the whole system and several safety criteria need to be considered for each component of the system.
Despite the importance of this issue, the idea of the multiple CBFs scheme is addressed in a few studies. In \cite{Rauscher2016}, multivariate CBFs of relative degree two are designed to safely control a manipulator. However, the idea is not extendable to safety constraints with arbitrary high relative degrees. Multiple safety constraints with relative degree one are also used in some applications such as obstacle avoidance in manipulator \cite{Landi2019}, constrained motion planning \cite{Saveriano2019}, and trajectory planning under temporal specifications \cite{Garg2019}. In this paper, we relax the assumption of relative degree one safety constraints for multiple CBFs structure.

Tractor-trailer systems are widely used in many applications ranging from road shipment to agricultural machinery. Although they are the most common mode of transportation in the supply chain, the poor maneuverability restricts their usage in safety-critical applications. Autonomous tractor-trailer robots (TTRs) can be a good solution to address this issue. TTRs generally consist of a car-like tractor which tows a passive or active trailer. Based on the trailer articulation structure, TTRs are divided into single-steering tractor-trailer robots (SSTTRs) and multi-steering tractor-trailer robots (MSTTRs). In particular, MSTTR has an additional input to navigate the trailer. The trailer has a rear-wheel independent steering system which gives a decent range of movements to the whole system. In consequence, MSTTRs have improved mobility compared to SSTTRs.


An accurate tracking control method is necessary for TTRs' autonomy but not sufficient. An obstacle avoidance control technique which guarantees safety, is essential for autonomous transportation. In the literature, obstacle avoidance of TTRs is mainly addressed in path planning methods. A multi-layered obstacle avoidance path planning approach for MSTTRs based on the particle swarm optimization method has been developed in \cite{Yuan2018}. In \cite{Zhao2020}, an online trajectory planning method has been developed for the dynamic model of a tractor with an arbitrary number of trailers. In \cite{Lashkari2019}, collision-free paths were generated via a cooperative approach (using workspace information). As an alternative technique, obstacle avoidance was guaranteed via a reactive method (using installed sensors on the robot). In this work, we aim to develop a tracking control strategy enhanced with a reactive obstacle avoidance technique, which effectively uses the idea of a multiple CBFs scheme to generate a safe motion for both the tractor and trailer in the presence of several obstacles. To the best of our knowledge, this study is the first attempt to use CBFs to synthesize controllers enforcing safety over tractor-trailer systems.

In this paper, we extend the results of the CBFs method to a multiple CBFs scheme where a safety constraint is designed for the tractor and trailer separately. Due to the successful applications of MPC to TTRs, linear time-varying model predictive control (LTV MPC) is selected as the nominal tracking control. The main contributions of this paper with respect to prior work are presented as follows.
\begin{itemize}
\item We introduce the multiple CBFs structure to design several safety constraints with high relative degrees. The systematic approach enables us to extend the existing formal construction method of a single CBF to multiple CBFs of arbitrary relative degrees.
\item We establish the multiple CBFs structure and ensure the forward invariance of the associated safe sets. Also, the Lipschitz continuity of the resulting safe control law is addressed.
\item The proposed method has been used to design a reactive obstacle avoidance technique for a class of tractor-trailer systems that ensures collision-free maneuver for both the tractor and trailer.
\end{itemize}

This paper is organized as follows.  The preliminary background on CBF method is provided in Section \ref{sec2}. The notion of multiple CBFs is introduced in Section \ref{sec3}, and an application of the proposed method is developed in Section \ref{sec4}. Simulation results are presented in Section \ref{sec5}. Finally, Section \ref{sec6} summarizes the conclusions of the paper.

\section{BACKGROUND AND PRELIMINARIES}\label{sec2}
Consider the input-affine control system
\begin{equation}
    \dot {\boldsymbol{x}} = f(\boldsymbol{x})+ G(\boldsymbol{x}) \boldsymbol{u},
    \label{eq1}
\end{equation}where state $\boldsymbol{x} \in \mathbb{R}^n$, input $\boldsymbol{u} \in \mathcal{U} \subset \mathbb{R}^m$, where $\mathcal{U}$ is a compact and convex set. Also, $G(\boldsymbol{x}) = [g_1(\boldsymbol{x}), g_2(\boldsymbol{x}), \dots, g_m(\boldsymbol{x})]$ and the vector fields $f:\mathbb{R}^n \rightarrow \mathbb{R}^n$ and $g_i:\mathbb{R}^n \rightarrow \mathbb{R}^{n}$, $i \in \{1, \dots, m \}$, are locally Lipschitz in $\boldsymbol{x}$. Let $\boldsymbol{u}=\pi(\boldsymbol{x})$ be a feedback control law such that the closed-loop system $\dot {\boldsymbol{x}} = F_{cl}(\boldsymbol{x})$ is locally Lipschitz. Then, for any initial state $\boldsymbol{x}(0) = \boldsymbol{x}_0$ and any input $\boldsymbol{u}(\cdot)$ that is locally bounded, the system (\ref{eq1}) has a unique solution $\boldsymbol{x}(t)$ defined on a maximal interval of existence $I(\boldsymbol{x}_0) = [0, I_{max})$. Next, we define the notion of forward invariance of a given set.
\begin{defn}\label{def1}
	The set $\Omega \subset \mathbb{R}^n$ is said to be forward invariant for a system of the form $\dot{\boldsymbol{x}} = F_{cl}(\boldsymbol{x})$, if for all $x(0) \in \Omega$, the solution $\boldsymbol{x}(t) \in \Omega$ for $t \geq 0$.
\end{defn}

If there exists a set of all safe states, forward invariance of that set implies the safety of the system. We use CBFs to certify the safety of the control system (\ref{eq1}). CBFs effectively use the idea of Nagumo's theorem to ensure the forward invariance of the desired set. Consider the safe set $\mathcal{C}$ as
\begin{align}
    \mathcal{C}={}& \{\boldsymbol{x} \in \mathbb{R}^n|h(\boldsymbol{x})\geq 0\},\label{eq2}\\
    \partial \mathcal{C}={}& \{\boldsymbol{x} \in \mathbb{R}^n|h(\boldsymbol{x})=0\},\label{eq3}\\
    Int(\mathcal{C})={}& \{\boldsymbol{x} \in \mathbb{R}^n|h(\boldsymbol{x})>0\},\label{eq4}
\end{align}where $h:\mathbb{R}^n \rightarrow \mathbb{R}$ is a continuously differentiable function and is designed such that the system satisfies a safety criterion. We assume that the initial state $\boldsymbol{x}_0 \in \mathcal{C}$, and ensure that the solution will never leave this set.
\begin{defn}\label{def2}
    (Control barrier function, \cite{ames2019}) Given the set $\mathcal{C}$ as defined in (\ref{eq2}), the continuously differentiable function $h(\boldsymbol{x})$ is called a control barrier function on a domain $D$ with $\mathcal{C} \subseteq D \subset \mathbb{R}^n$, if there exists a class $\mathcal{K}$ function\footnote{A continuous function $\alpha:[0, a) \rightarrow [0, \infty$), is said to belong to class $\mathcal{K}$ if it is strictly increasing and $\alpha(0)=0$.} $\alpha$ such that 
\begin{equation}
    \sup_{\boldsymbol{u} \in \mathcal{U}} [L_f h(\boldsymbol{x})+\sum^m_{i=1} L_{g_i} h(\boldsymbol{x})u_i] \geq -\alpha(h(\boldsymbol{x})),
    \label{eq5}
\end{equation}for all $\boldsymbol{x} \in D$, where $L_f{h(\boldsymbol{x})}$ and $L_{g_i}{h(\boldsymbol{x})}$ are the Lie derivatives of $h(\boldsymbol{x})$ with respect to the vector fields $f$ and $g_i$, and $u_i$ are the elements of the input vector $\boldsymbol{u}$.
\end{defn}
The inequality (\ref{eq5}) provides a condition for obtaining a control law. Given a CBF $h(\boldsymbol{x})$, define the following set for all $\boldsymbol{x} \in D$ as
\begin{equation}
        \mathcal{U}_{s}(\boldsymbol{x}) \triangleq \{\boldsymbol{u} \in \mathcal{U} | L_f h(\boldsymbol{x})+\sum^m_{i=1} L_{g_i} h(\boldsymbol{x})u_i +\alpha(h(\boldsymbol{x})) \geq 0 \},
        \nonumber
\end{equation}
which contains the controls that render the set $\mathcal{C}$ forward invariant for system (\ref{eq1}). We have the following Corollary form \cite{ames2016control} for the forward invariance of the set $\mathcal{C}$.
\begin{cor}\label{cor1}
Consider the CBF $h(\boldsymbol{x}):\mathbb{R}^n \rightarrow \mathbb{R}$ with the associated set $\mathcal{C}$ in (\ref{eq2}), any Lipschitz continuous controller $\boldsymbol{u}(\boldsymbol{x}) \in \mathcal{U}_{s}(\boldsymbol{x})$, guarantees that the set $\mathcal{C}$ is froward invariant for the system (\ref{eq1}).
\end{cor}

Note that Definition \ref{def2} is only applicable to systems with relative degree one. However, in many applications, CBFs have relative degrees greater than one. A modification of the original theory is investigated in \cite{nguyen2016exponential}, which introduced the exponential control barrier functions (ECBFs). Also, a more general definition of high-order control barrier functions (HOCBF) has been introduced in \cite{xiao2021high}. These works extended the results to a single CBF with a higher relative degree than one. In the next section, we deploy the idea of HOCBF to verify the forward invariance for the case that we have multiple CBFs with high relative degrees.
\section{MULTIPLE CONTROL BARRIER FUNCTIONS}\label{sec3}
We begin with defining the safe sets associated with candidate CBFs with the goal to design a controller that guarantees the forward invariance of these sets
\begin{equation}
    \mathcal{C}_i= \{\boldsymbol{x} \in \mathbb{R}^n|h_i(\boldsymbol{x})\geq 0\},\label{eq6}
\end{equation}where $h_i:\mathbb{R}^n \rightarrow \mathbb{R}$, $i \in \{1, \dots, m\}$, are continuously differentiable functions that we design to satisfy several safety criteria. The sets $\partial \mathcal{C}_i$ and $Int(\mathcal{C}_i)$ will be defined similar to (\ref{eq3}) and (\ref{eq4}) for each $h_i(\boldsymbol{x})$, respectively.
We define the set of admissible states as
\begin{equation}
\mathcal{A} = \bigcap^m_{i=1} \mathcal{C}_i = \{\boldsymbol{x} \in \mathbb{R}^n|\forall 1\leq i \leq m, h_i(\boldsymbol{x}) \geq 0 \}.
\label{eq7}
\end{equation}We want to verify the forward invariance for all sets $\mathcal{C}_i$, which is equivalent to the forward invariance of their intersection. Therefore, we assume that the set of admissible states $\mathcal{A}$ is non-empty.

For a given $i \in \{1, \dots, m\}$, suppose $h_i(\boldsymbol{x})$ has an arbitrary relative degree $r_i$, which is defined to be the smallest integer such that at least one of the inputs explicitly appears in the time derivatives of $h_i(\boldsymbol{x})$:
\begin{equation}
    h_i^{(r_i)}(\boldsymbol{x}) = L_f^{r_i} h_i(\boldsymbol{x}) + \sum^m_{l=1}(L_{g_l}L_f^{r_i - 1} h_i(\boldsymbol{x}))u_l,
    \label{eq8}
\end{equation}where at least one $L_{g_l}L_f^{r_i - 1} h_i(\boldsymbol{x})$ is non-zero for all $\boldsymbol{x} \in D \subseteq \mathbb{R}^n$. If we repeat this procedure for all $i \in \{1, \dots, m \}$, equation  (\ref{eq8}) can be written in the matrix form 
\begin{equation}
    \begin{bmatrix}
        h_1^{(r_1)}\\
        \vdots\\
        h_m^{(r_m)}
    \end{bmatrix} = 
    \begin{bmatrix}
        L_f^{r_1}h_1\\
        \vdots\\
        L_f^{r_m}h_m
    \end{bmatrix} + E(\boldsymbol{x})
    \begin{bmatrix}
        u_1\\
        \vdots\\
        u_m
    \end{bmatrix},
    \label{eq9}
\end{equation}where $h_i(\boldsymbol{x})$ is denoted by $h_i$ for short. The matrix $E(\boldsymbol{x})$ is obtained by
\begin{equation}
    E(\boldsymbol{x}) =
    \begin{bmatrix}
        L_{g_1}L_f^{r_1 - 1} h_1 & \ldots & L_{g_m}L_f^{r_1 - 1} h_1\\
        \vdots &  \ddots & \vdots\\
        L_{g_1}L_f^{r_m - 1} h_m & \ldots & L_{g_m}L_f^{r_m - 1} h_m
    \end{bmatrix},
    \label{eq10}
\end{equation} which is called the decoupling matrix.
\begin{defn}\label{def3}
The system (\ref{eq1}) is said to have a vector relative degree $\{ r_1, r_2, \dots, r_m \}$ if $L_{g_l}L_{f}^k h_i(\boldsymbol{x}) = 0$, for all $1 \leq l \leq m$, for all $1 \leq i \leq m$, for all $k < r_i-1$, and the decoupling matrix $E(\boldsymbol{x})$ is non-singular for all $\boldsymbol{x} \in D$.
\end{defn}

We define the notion of multiple CBFs with high relative degrees as multi-CBFs and prove the forward invariance of the safe set associated with them. Toward this end, the following definition and results will be helpful.

\begin{lem}\label{lem1}
Let $\alpha$ be a locally Lipschitz class $\mathcal{K}$ function and $h:[0,t_f) \rightarrow \mathbb{R}$ ($t_f$ could be infinity) be a continuous function. If $\dot h(t) \geq -\alpha(h(t))$ for all $t \in [0,t_f)$, and $h(0)\geq0$, then there exists a class $\mathcal{KL}$ function\footnote{A continuous function $\beta:[0,a) \times [0,\infty) \rightarrow [0,\infty)$ is said to belong to class $\mathcal{KL}$, if for each fixed $s$, the mapping $\beta(r,s)$ belongs to class $\mathcal{K}$ functions with respect to $r$, and for each fixed $r$, $\beta(r,s)$ is decreasing with respect to $s$ and $\beta(r,s) \rightarrow 0$ as $s \rightarrow \infty$.} $\beta:[0,\infty)\times[0,\infty) \rightarrow [0,\infty)$ such that $h(t)\geq\beta(h(0),t)$, and $h(t)\geq0$, for all $t \in [0,t_f)$.
\end{lem}
\begin{proof}
We refer to \cite{glotfelter2017nonsmooth} for a proof based on the Lemma 4.4 in \cite{khalil2002nonlinear} and comparison techniques.
\end{proof}

In order to ensure the forward invariance of each safe set $\mathcal{C}_i$ in (\ref{eq6}), we define the set of functions $m^j_{i}:\mathbb{R}^n \rightarrow \mathbb{R}$ as
\begin{equation}
m^j_{i}(\boldsymbol{x}) = \dot m^{j-1}_{i}(\boldsymbol{x}) + \alpha^j_i(m^{j-1}_{i}(\boldsymbol{x})),
\label{eq17}
\end{equation}where $\alpha^j_i$ is a class $\mathcal{K}$ function with $m^0_i(\boldsymbol{x})=h_i(\boldsymbol{x)}$ for $j \in \{1, \dots, r_i \}$, and $i \in \{1, \dots, m \}$. We also define their super level sets $\mathcal{M}^j_i$ and the interior of these sets as
\begin{align}
\mathcal{M}^j_i ={}& \{\boldsymbol{x} \in \mathbb{R}^n|m^{j-1}_i(\boldsymbol{x}) \geq 0\},\label{eq18}\\
Int(\mathcal{M}^j_i) ={}& \{\boldsymbol{x} \in \mathbb{R}^n|m^{j-1}_i(\boldsymbol{x}) > 0\},
\label{eq19}
\end{align}
for all $j \in \{1, \dots, r_i \}$, and $i \in \{1, \dots, m \}$.
\begin{defn}\label{def4}
(Multi-CBFs) Let the functions $m^j_i(\boldsymbol{x})$ and sets $Int(\mathcal{M}^j_i)$ be defined by (\ref{eq17}) and (\ref{eq19}), respectively. The set of $m$ continuously differentiable functions $\{ h_1, \dots, h_m \}$, where $h_i:\mathbb{R}^n \rightarrow \mathbb{R}$, $i \in \{1, \dots, m \}$, is called a multi-CBFs set if $h_i$ and its derivatives up to order $r_i$, are locally Lipschitz continuous, and there exists a set of continuously differentiable class $\mathcal{K}$ functions $\alpha^j_i$,  $j \in \{1, \dots, r_i \}$, such that $m^{r_i}_i(\boldsymbol{x}) \geq 0$. This inequality can be represented as
\begin{align}\label{eq20}
\sup_{u \in \mathcal{U}}[L_f^{r_i}h_i(\boldsymbol{x})+ \sum^m_{l=1}(L_{g_l}L_f^{r_i - 1} h_i(\boldsymbol{x}))u_l+R(h_i)+\nonumber\\
\alpha^{r_i}_i(m^{r_i-1}_i(\boldsymbol{x}))] \geq 0,
\end{align}for each $i \in \{ 1, \dots, m \}$, for all $\boldsymbol{x} \in \bigcap^{m}_{i=1} \bigcap^{r_i}_{j=1} Int(\mathcal{M}^j_i)$, where $R(h_i)$ denotes the Lie derivatives of $h_i$ with respect to $f$ up to order $r_i-1$.
\end{defn}
Based on Definition \ref{def4}, we state the main theorem of this paper for the forward invariance of the associated sets.
\begin{thm}\label{thm2}
Given system (\ref{eq1}) with a set of multi-CBFs $\{h_1, \dots, h_m \}$, any Lipschitz continuous control $\boldsymbol{u}$ that satisfies $m$ constraints of the form (\ref{eq20}) renders the set $\bigcap^{m}_{i=1} \bigcap^{r_i}_{j=1} Int(\mathcal{M}^j_i)$ forward invariant for system (\ref{eq1}).
\end{thm}
\begin{proof}
Since $\{h_1, \dots, h_m \}$ is a set of multi-CBFs, the derivatives of $h_i$ are locally Lipschitz. Also, the continuous differentiability of $\alpha^j_i$ implies their Lipschitz continuity. Thus, $m^j_i(\boldsymbol{x})$, $j\in\{1,\dots, r_i-1 \}$ are Lipschitz continuous. The control $\boldsymbol{u}$ appears in the $r_i^{th}$ time derivative of $h_i$, thus it only exists in the expression of $m^{r_i}_i(\boldsymbol{x})$. Since $\boldsymbol{u}$ is also Lipschitz continuous, $m^{r_i}_i(\boldsymbol{x})$ is Lipschitz continuous.
Since $h_i(\boldsymbol{x})$ belongs to a multi-CBFs set, then $m^{r_i}_i(\boldsymbol{x}) \geq 0$, i.e., $\dot{m}^{r_i-1}_i(\boldsymbol{x}) + \alpha^{r_i}_i(m^{r_i-1}_i(\boldsymbol{x})) \geq 0$, for all $\boldsymbol{x} \in \bigcap^{r_i}_{j=1}\mathcal{M}^j_i$. By Lemma \ref{lem1}, since $\boldsymbol{x}_0 \in \mathcal{M}^{r_i}_i$, $m^{r_i - 1}_i(\boldsymbol{x}) > 0$, which implies that $\boldsymbol{x}(t) \in Int(\mathcal{M}^{r_i}_i)$ for all $t \geq 0$. In consequence, we have $\dot{m}^{r_i-2}_i(\boldsymbol{x}) + \alpha^{r_i-1}_i(m^{r_i-2}_i(\boldsymbol{x})) > 0$. We use Lemma \ref{lem1} another time. Since $\boldsymbol{x}_0 \in Int(\mathcal{M}^{r_i-1}_i)$, we also have  $m^{r_i - 2}_i(\boldsymbol{x}) > 0$, which implies that $\boldsymbol{x}(t) \in Int(\mathcal{M}^{r_i -1}_i)$ for all $t \geq 0$. Repeating this procedure, we get that $\boldsymbol{x}(t) \in Int(\mathcal{M}^j_i)$, for all $j \in \{ 1, \dots, r_i \}$, for all $t \geq 0$. Therefore, the set $\bigcap^{m}_{i=1} \bigcap^{r_i}_{j=1} Int(\mathcal{M}^j_i)$  is forward invariant for system (\ref{eq1}).
\end{proof}
We have addressed the case that the number of inputs and CBFs are the same (i.e., the decoupling matrix is square). The following remark generalizes the idea to systems with a different number of inputs and CBFs. 
\begin{rem}\label{rem3}
The results established so far can be extended to a system having a different number of inputs and CBFs. Note that in order to have vector relative degree, the decoupling matrix must have full row rank (equal to the number of CBFs).
\end{rem}
\begin{rem}\label{rem4}
If the decoupling matrix is singular, the system does not have a vector relative degree. In this case, we can use dynamic extension, which is essentially choosing new inputs as the derivative of the existing system inputs such that the corresponding decoupling matrix becomes nonsingular \cite{isidori1995nonlinear}.  The control system is designed based on the additional dynamics and the new set of inputs, and the original inputs will be computed by integration.
\end{rem}
\subsection{Quadratic programming optimization}
In this section, we show that the multiple CBFs structure can be combined with an arbitrary nominal controller to enforce safety. Suppose we have a nominal controller designed to satisfy only the control objectives, and it does not necessarily ensure the system's safety. We want this control law to be applied to the system only if it does not violate safety criteria. Otherwise, we need to rectify it such that the resulting safe input has the least deviation from the nominal control law. We have obtained a set of $m$ safety constraints of the form (\ref{eq20}), which are affine in input $\boldsymbol{u}$. The following online QP optimization rectifies the nominal control law to generate a safe input
\begin{align}
    &\boldsymbol{u}_{s}(\boldsymbol{x}) = \underset{\boldsymbol{u} \in \mathcal{R}^m}{\arg\min} \hspace{4pt} {\boldsymbol{u}^T\boldsymbol{u}-2\boldsymbol{u}_{nom}^T\boldsymbol{u}} \label{eq21}\\
    &\textrm{s.t.} \quad A_{mcbf}(\boldsymbol{x}) \boldsymbol{u} \leq b_{mcbf}(\boldsymbol{x}),\label{eq22}
\end{align}
where $\boldsymbol{u}_{nom}$ is a locally Lipschitz continuous nominal control law. The constraint (\ref{eq22}) is the matrix form of (\ref{eq20}), where
\begin{align}
     &A_{mcbf}(\boldsymbol{x}) = -E(\boldsymbol{x}),\nonumber\\
     &b_{mcbf}(\boldsymbol{x}) =\quad \begin{bmatrix}
        L_f^{r_1}h_1 +R(h_1)+\alpha^{r_1}_1(m^{r_1-1}_1(\boldsymbol{x}))\\
        \vdots\\
        L_f^{r_m}h_m +R(h_m)+\alpha^{r_m}_m(m^{r_m-1}_m(\boldsymbol{x}))
    \end{bmatrix}.
    \label{eq23_2}
\end{align}
Note that the nonsingularity of the decoupling matrix $E(\boldsymbol{x})$ is essential for finding a safe control input $\boldsymbol{u}_s(\boldsymbol{x})$ in (\ref{eq21}). The obtained control law $\boldsymbol{u}_s$ must be Lipschitz continuous as it is one of the fundamental requirements of the Theorem \ref{thm2}. In the following theorem, the Lipschitz continuity of the solution is investigated.
\begin{thm}\label{thm3}
Consider the system (\ref{eq1}) with a set of multi-CBFs $h_i:\mathbb{R}^n \rightarrow \mathbb{R}$, $i\in\{1,\dots,m \}$, and a locally Lipschitz nominal control law $\boldsymbol{u}_{nom}$. Let the set of admissible states $\mathcal{A}$ be defined as (\ref{eq7}). Then the control law $\boldsymbol{u}_s$ obtained by solving the QP optimization (\ref{eq21}) is Lipschitz continuous.
\end{thm}
\begin{proof}
Suppose the $\boldsymbol{u}_s(\boldsymbol{x})$ is the solution of (\ref{eq21}). Based on the Mangasarian-Fromovitz regularity conditions in \cite{mangasarian1967fritz} and its application to sufficient conditions for Lipschitz continuity of QP control in \cite{morris2013sufficient}, if the conditions
\begin{enumerate}[(i)]
\item $p^*(\boldsymbol{x}) > 0$, where $p^*(\boldsymbol{x})$ is the solution for the linear program
\begin{align}
    &p^*(\boldsymbol{x}) = \max_{(\boldsymbol{u}, p) \in \mathbb{R}^{m+1}} p \label{eq25_2}\\
    &\textrm{s.t.} \quad \begin{bmatrix}
        A_{mcbf}(\boldsymbol{x}) & \boldsymbol{1}_{m \times 1}
    \end{bmatrix} \begin{bmatrix}
        \boldsymbol{u}\\
        p
    \end{bmatrix}\leq b_{mcbf}(\boldsymbol{x})\label{eq26_2},
\end{align}
\item $A_{mcbf}(\boldsymbol{x})$ and $b_{mcbf}(\boldsymbol{x})$ are Lipschitz continuous at $\boldsymbol{x}$,
\item The nominal control law $\boldsymbol{u}_{nom}(\boldsymbol{x})$ is Lipschitz continuous at $\boldsymbol{x}$,
\end{enumerate}
hold at a state $\boldsymbol{x}\in \mathbb{R}^n$, then the solution $\boldsymbol{u}_s(\boldsymbol{x})$ is unique and Lipschitz at $\boldsymbol{x}$. The uniqueness of the solution is based on the results in \cite{mangasarian1967fritz} and is verified by (i). We can rewrite the linear program (\ref{eq25_2}) as
\begin{equation*}
    p^*(\boldsymbol{x}) = \max_{\substack{\boldsymbol{u} \in \mathbb{R}^m \\ 1 \leq i \leq m}} L_f^{r_i}h_i +R(h_i)+\alpha^{r_i}_i(m^{r_i-1}_i(\boldsymbol{x})) + L_G h_i(\boldsymbol{x}) \boldsymbol{u},
\end{equation*}
where $L_G h_i(\boldsymbol{x}) = [L_{g_1}L_f^{r_i-1}h_i(\boldsymbol{x}) \dots L_{g_m}L_f^{r_i-1}h_i(\boldsymbol{x})]$ is a row vector. The derivation above is possible since the set $\mathcal{A}$ is assumed to be  non-empty, which implies that the constraints do not conflict. Now, we can consider the linear program 
\begin{equation*}
    p^*(\boldsymbol{x}) = \max_{\boldsymbol{u} \in \mathbb{R}^m} p(\boldsymbol{x}),
\end{equation*}
with the equality constraints $p(\boldsymbol{x}) = L_f^{r_i}h_i +R(h_i)+\alpha^{r_i}_i(m^{r_i-1}_i(\boldsymbol{x})) + L_G h_i(\boldsymbol{x}) \boldsymbol{u}$, for all $i \in \{1, \dots, m \}$. As the multi-CBFs satisfies (\ref{eq20}), $p(\boldsymbol{x}) \geq 0$. By Theorem \ref{thm2}, starting from the interior of the set of admissible states $\mathcal{A}$, the trajectory remains in the interior set. Thus, $p(\boldsymbol{x})>0$ is fulfilled on $\mathcal{A}$, verifies condition (i). Condition (ii) holds since $\alpha^j_i$, $h_i(\boldsymbol{x})$ and its derivatives are locally Lipschitz (based on Definition \ref{def4}), which implies the Lipschitz continuity of $A_{mcbf}(\boldsymbol{x})$ and $b_{mcbf}(\boldsymbol{x})$. Condition (iii) also holds as the local Lipschitz continuity of the nominal control law $\boldsymbol{u}_{nom}$ has been assumed, which completes the proof. 
\end{proof}

\section{AN APPLICATION TO AUTONOMOUS TRACTOR-TRAILER SYSTEMS}\label{sec4}
In this section, the problem of obstacle avoidance while tracking a given reference trajectory is addressed for autonomous tractor-trailer systems. The reference trajectory is not necessarily safe, and we assume that there might be some obstacles on the reference trajectory. Thus, reactive obstacle avoidance based on CBFs is required. The structure of the MSTTR enables us to deploy multi-CBFs method. The trajectory tracking objective is achieved using an LTV MPC approach as the nominal controller, and a multiple CBFs scheme is designed for the tractor and trailer to ensure the system's safety.
\subsection{Kinematic model of the tractor-trailer robot}
\begin{figure}[htbp]
\centerline{\includegraphics[trim=0.2cm 0.7cm 0.2cm 0.2cm, clip, totalheight=0.2\textheight]{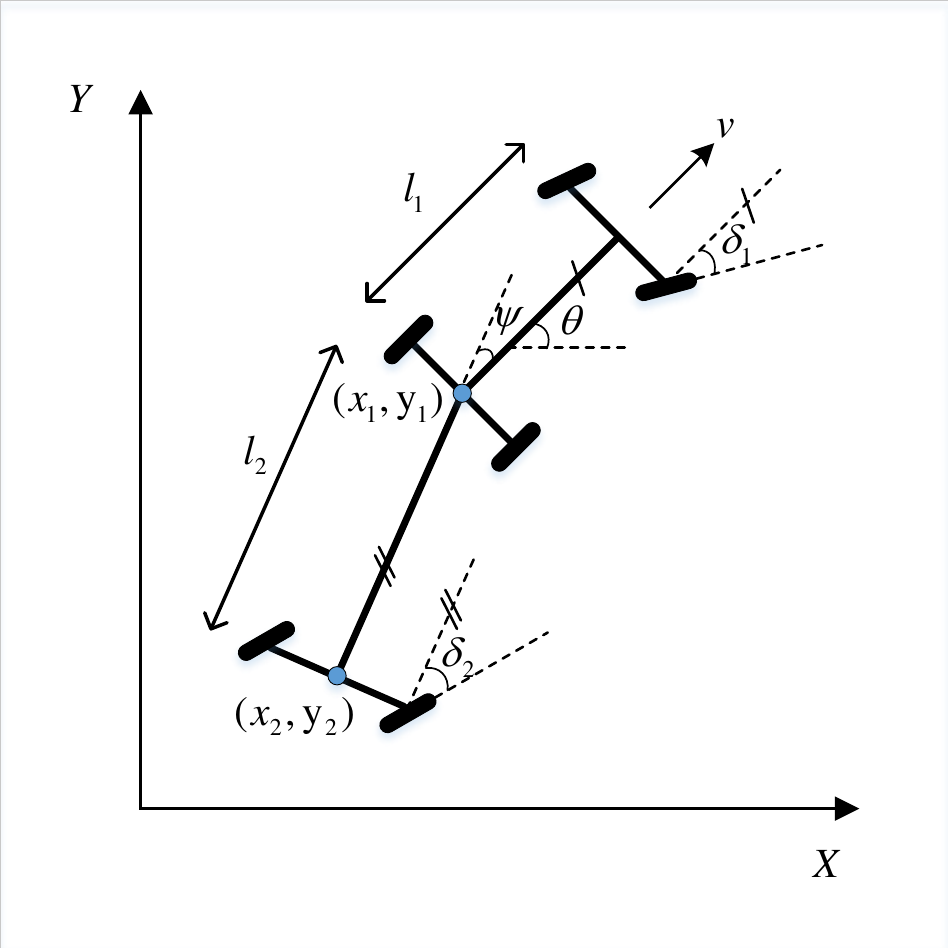}}
\caption{A multi-steering tractor-tractor system.}
\label{fig1}
\end{figure}

The multi-steering tractor-trailer mobile robot comprises a car-like tractor and a trailer equipped with actively steerable wheels. Each body has a reference point located on the midpoint of its rear axle. Let $p_i = [x_i, y_i]^T, i=1,2$, be the coordinates of the reference points on the Cartesian plane. The steering angles of the tractor and trailer are $\delta_1$ and $\delta_2$, respectively. The angles $\theta$ and $\psi$ denote the orientation of the tractor and the relative angle between two vehicle segments, respectively. $v$ is the linear velocity, and $a$ is the acceleration of the tractor. We consider low-speed turning maneuvers in this work. Thus, the assumption of pure rolling of the wheels holds and wheel slip dynamical effects can be neglected. Figure \ref{fig1} shows an on-axle tractor-trailer mobile robot. The state vector of the system denotes by $\boldsymbol{x}=[x_1, y_1, v, a, \theta, \psi, \delta_1, \delta_2]^T$ and the kinematic model of the system can be described by
\begin{equation}
    \begin{aligned}
        & \dot x_1 = v \cos \theta,\\
        & \dot y_1 = v \sin \theta,\\
        & \dot v = a,\\
        & \dot a = J,\\
        & \dot \theta = \frac{v}{l_1} \tan \delta_1,\\
        & \dot\psi = \frac{v}{l_1} \tan \delta_1 - \frac{v}{l_2} (\tan \delta_2 \cos \psi + \sin \psi),\\
        & \dot \delta_1 = \omega_1,\\
        & \dot \delta_2 = \omega_2,
        \label{eq23}
    \end{aligned}
\end{equation}
where $l_1$ and $l_2$ are the distances between the front and rear axle (wheelbase) of the bodies. The coordinates of the trailer's reference point are given by
\begin{equation}
    \begin{aligned}
        & x_2 = x_1 - l_2 \cos (\theta - \psi),\\
        & y_2 = y_1 - l_2 \sin(\theta - \psi).
    \end{aligned}
\label{eq24}
\end{equation}

The control input is represented by $\boldsymbol{u} = [u_1, u_2, u_3]^T =[J, \omega_1, \omega_2]^T$, where $J$ is the rate of change of the acceleration and $\omega_1$ is the angular velocity of the tractor, and $\omega_2$ is the angular velocity of the trailer.
\subsection{Construction of CBFs}
Position-based barrier functions are used to construct safe sets for both the tractor and trailer to ensure collision-free maneuvers for the whole system. 
Consider we know the exact position of the robot at each time step. The position of the obstacles are denoted by $p^o_k=[x_k^o, y_k^o]^T, k\in \{1,\dots,N_o\}$, where $N_o$ is the number of obstacles in the workspace. We assume that obstacles are distributed in the workspace such that there exists a collision-free path around the reference trajectory.

We use multiple CBFs scheme to construct the CBFs for this system. The CBFs $h_1(\boldsymbol{x})$ and $h_2(\boldsymbol{x})$ are the CBFs associated with the tractor and trailer, respectively. Suppose that we have $N_o$ obstacles in the workspace, and for each of them, we design $h_{ik}(\boldsymbol{x})$, $i=1,2$, $k \in \{1, \dots, N_o \}$. These CBFs are obtained based on the position of the bodies' reference points $p_i$, and the obstacles positions $p_k^o$. The safe sets and their associated CBFs for the tractor and trailer are given by
\begin{align}
    &\mathcal{C}_{ik} = \{\boldsymbol{x} \in \mathbb{R}^n|h_{ik}(\boldsymbol{x}) \geq 0 \}, \label{eq25}\\
    & h_{ik}(\boldsymbol{x})= (x_i - x^o_k)^2 + (y_i - y^o_k)^2 - {d_i}^2, \label{eq26}
\end{align}for all $i=1,2$, $k \in \{1, \dots, N_o\}$, where $d_i$ is a design parameter and denotes the safety distance for each vehicle segment. The CBFs $h_{1k}(\boldsymbol{x})$ and $h_{2k}(\boldsymbol{x})$ have relative degrees three and two, respectively, for all $k \in \{1, \dots,N_o \}$.

We choose linear functions with positive coefficients $k^j_i$ as the set of locally Lipschitz class $\mathcal{K}$ functions $\alpha^j_i(\boldsymbol{x})$, $i=1,2,j=\{1,\dots,r_i\}$, which are defined the same for all obstacles $k \in \{ 1, \dots, N_o \}$. Therefore, the constraints in (\ref{eq20}) can be identically written as
\begin{align}
      &\dddot h_{1k}(\boldsymbol{x},\boldsymbol{u})+ K_1  [h_{1k}(\boldsymbol{x}),\dot h_{1k}(\boldsymbol{x}), \ddot h_{1k}(\boldsymbol{x})]^T \geq 0,\label{eq27}\\
      & \ddot h_{2k}(\boldsymbol{x},\boldsymbol{u})+ K_2  [h_{2k}(\boldsymbol{x}),\dot h_{2k}(\boldsymbol{x})]^T \geq 0,\label{eq28}
\end{align}where $K_1=[k_1^0, k_1^1, k_1^2]$, $K_2=[k_2^0,k_2^1]$. Also, $\dddot h_{1k}(\boldsymbol{x},\boldsymbol{u})$ and $\ddot h_{2k}(\boldsymbol{x},\boldsymbol{u})$ with $k=\{1,\dots, N_o\}$ are affine in $\boldsymbol{u}$. We refer to the Appendix for the derivation of (\ref{eq27}) and (\ref{eq28}).
\subsection{Controller design}
This section addresses the tracking control design problem of the multi-steering tractor-trailer system with multiple CBFs. An LTV MPC approach is utilized for trajectory tracking. Thus, we first derive a linear approximation of the system and then discretize the obtained linearized time-varying system. Next, we formulate the discrete LTV-MPC controller, which focuses on the tracking of the reference trajectory. Then, we enforce the safety constraints (\ref{eq27}) and (\ref{eq28}) using the QP optimization introduced in (\ref{eq21}).
\subsubsection{Discrete-time LTV model of the kinematic model}
 In this section, we want to approximate the kinematic model of a tractor-trailer system to a discrete-time LTV model by online-linearization. Consider the nonlinear model in (\ref{eq23}), which can be represented as a generic nonlinear model $\dot x = F(\boldsymbol{x}, \boldsymbol{u})$. Suppose that we are given the reference state trajectory $\boldsymbol{x}^r(t)$, which is obtained by applying a reference input trajectory $\boldsymbol{u}^r(t)$ (given by a planner) to the system (\ref{eq23}). 
 We first obtain the linearized model using Taylor expansion around $(\boldsymbol{x}^r, \boldsymbol{u}^r)$, and then descritize the linearized system with the sampling time $T_s$ to obtain
 \begin{align}
    \boldsymbol{x}_{k+1} ={}& A_k \boldsymbol{x}_k + B_k \boldsymbol{u}_k + C_k,\label{eq26_3}\\
    A_k ={}& I + \frac{\partial F}{\partial \boldsymbol{x}}(\boldsymbol{x}^r, \boldsymbol{u}^r)T_s,\nonumber\\
    B_k ={}& \frac{\partial F}{\partial \boldsymbol{u}}(\boldsymbol{x}^r, \boldsymbol{u}^r)T_s,\nonumber\\
    C_k ={}& (F(\boldsymbol{x}^r, \boldsymbol{u}^r) - \frac{\partial F}{\partial \boldsymbol{x}}(\boldsymbol{x}^r, \boldsymbol{u}^r)\boldsymbol{x}^r - \frac{\partial F}{\partial \boldsymbol{u}}(\boldsymbol{x}^r, \boldsymbol{u}^r)\boldsymbol{u}^r)T_s.\nonumber
\end{align}
We refer to the Appendix, for a derivation of the system matrices in (\ref{eq26_3}).

\subsubsection{Discrete-time LTV MPC formulation}
Consider we have a full measurement of the state $\boldsymbol{x}_k$ at current time step $k$. The following optimization problem formulates the discrete LTV MPC design:
\begin{align}
    \min_{X_t^{opt}, U_t^{opt}} \quad & ||\boldsymbol{x}^r_{t+N|t} - \boldsymbol{x}_{t+N|t}||^2_{P} + \nonumber\\
    \quad & \sum^{t+N-1}_{k=t} ||\boldsymbol{x}^r_{k|t} - \boldsymbol{x}_{k|t}||^2_Q + ||\boldsymbol{u}^r_{k|t} - \boldsymbol{u}_{k|t}||^2_R \label{eq31}\\
    \textrm{s.t.} \quad & \boldsymbol{x}_{k+1|t} = A_k \boldsymbol{x}_{k|t} + B_k \boldsymbol{u}_{k|t}+C_k, \nonumber\\
    \quad &  k=t, \dots, t+N-1, \nonumber\\
    \quad & \boldsymbol{x}_{t|t} = \boldsymbol{x}(t),\nonumber\\
    \quad & \boldsymbol{x}_{k|t} \in \mathcal{X}, \boldsymbol{u}_{k|t} \in \mathcal{U},\nonumber
 \end{align}
where $X^{opt}_t=[\boldsymbol{x}_{t|t}, \dots, \boldsymbol{x}_{t+N-1|t}]^T$ is the vector of optimized states, and $U^{opt}_t=[\boldsymbol{u}_{t|t}, \dots, \boldsymbol{u}_{t+N-1|t}]^T$ is the vector of optimized inputs in the prediction horizon $N$. The notation $\boldsymbol{x}_{t+k|t}$ is used to denote the state $\boldsymbol{x}$ at time step $t+k$ predicted at the current time $t$. The weighting matrices $P$ and $Q$ are positive semi definite and $R$ is positive definite and penalize the difference between the final state, state, and the input with their reference values, respectively. The nominal control law is denoted by $\boldsymbol{u}_{nom}$, which is obtained by picking the first element of $U^{opt}_t$.

\begin{table}[htbp]
\caption{Maximum values for the MSTTR and SSTTR models}
\label{tab1}
\centering
\begin{tabular}{|ll|ll|l}
\cline{1-4}
\multicolumn{2}{|l|}{MSTTR Parameters} & \multicolumn{2}{l|}{SSTTR Parameters} &  \\ \cline{1-4}
\multicolumn{1}{|l}{$v_{max}$} & $20 m/s$  & \multicolumn{1}{l}{$v_{max}$} & $20 m/s$ &  \\ 
\multicolumn{1}{|l}{$a_{max}$}  & $1 m/s^2$  & \multicolumn{1}{l}{$\delta_{1_{max}}$}  &  $0.784 rad$ &  \\ 
\multicolumn{1}{|l}{$\psi_{max}$}  &  $0.784 rad$ & \multicolumn{1}{l}{$a_{max}$}  & $1 m/s^2$ &  \\
\multicolumn{1}{|l}{$\delta_{1_{max}}$}  &  $0.784 rad$ & \multicolumn{1}{l}{$\omega_{1_{max}}$}  & $1.5 rad/s$ &  \\
\multicolumn{1}{|l}{$\delta_{2_{max}}$}  & $0.784 rad$  & \multicolumn{1}{l}{}  &  &  \\ 
\multicolumn{1}{|l}{$J_{max}$}  & $2.5 m/s^3$  & \multicolumn{1}{l}{}  &  &  \\ 
\multicolumn{1}{|l}{$\omega_{1_{max}}$}  & $1.5 rad/s$  & \multicolumn{1}{l}{}  &  &  \\ 
\multicolumn{1}{|l}{$\omega_{2_{max}}$}  &  $0.5 rad/s$ & \multicolumn{1}{l}{}  &  &  \\ \cline{1-4}
\end{tabular}
\end{table}

    
\section{SIMULATION RESULTS}\label{sec5}
In this section, simulation studies highlight the effectiveness of the proposed control method. A comparison is made between the multiple CBFs scheme and ECBF method. Then, the accuracy of the proposed control strategy is investigated in a cluttered environment. Simulations are run in the Python environment, and the optimization problems are solved using CVXPY \cite{Diamond2016} package.

Consider the system described in (\ref{eq23}), the tractor and trailer wheelbases are $l_1 = 2.5\hspace{2 pt}m$ and $l_2 = 5.5 \hspace{2 pt}m$, respectively. We have the compact sets
\begin{align}
    \mathcal{X}={}&\{\boldsymbol{x}\in \mathbb{R}^8:\hspace{2pt}|\boldsymbol{x}_i| \leq \boldsymbol{x}_{i_{max}} \}, i=1, \dots, 8, \label{eq40}\\
    \mathcal{U} ={}& \{ \boldsymbol{u} \in \mathbb{R}^3:\hspace{2pt}|\boldsymbol{u}_i| \leq \boldsymbol{u}_{i_{max}} \}, i=1, 2, 3,
    \label{eq41}
\end{align}
where $\boldsymbol{x}_i$ and $\boldsymbol{u}_i$ are the constrained elements of the state and input vectors, respectively. $\boldsymbol{x}_{i_{max}}$ and $\boldsymbol{u}_{i_{max}}$ are the maximum values which can be obtained from Table \ref{tab1}.

The multi-CBFs design parameters are safety distances $d_1 = 4.6 \hspace{2pt} m$, $d_2 = 3 \hspace{2pt} m$, and gain vectors $K_1 = [1, 2, 2]^T$, $K_2 = [4, 4]^T$. The LTV MPC parameters are the prediction horizon $N=5$, the sampling time $T_s =  0.2 \hspace{2pt} s$, the weighting matrices $Q=diag(1, 1, 0.5, 0, 0.05, 0.1, 0, 0)$, $P = Q$, and $R=diag(0.01, 0.05, 0.05)$.
\subsection{Obstacle avoidance in a $90^{\circ}$ turning maneuver}
The task is to follow a $90^{\circ}$ turning maneuver path while avoiding two obstacles. In this simulation, the performance of the multi-CBFs structure application to MSTTR has been compared with the application of ECBF method to the SSTTR (an SSTTR model can be obtained by substituting $\delta_2=0$ in (\ref{eq23})). As the position-based CBF has relative degree two with respect to SSTTR system, we have applied ECBF method \cite{nguyen2016exponential} to this system. The ECBF gain vector is $K'_1 = [1, 2]^T$ and the weighting matrices are $Q'=P'= diag(1, 1, 0.5, 0.5, 0.5)$, and $R'=diag(0.01, 0.05)$. Other parameters are the same in both cases. Identical parameters have been used in LTV MPC and CBF design for both systems.

\begin{figure}[htbp]
\centerline{\includegraphics[trim=0.2cm 8.0cm 0.2cm 8.0cm, clip, totalheight=0.15\textheight]{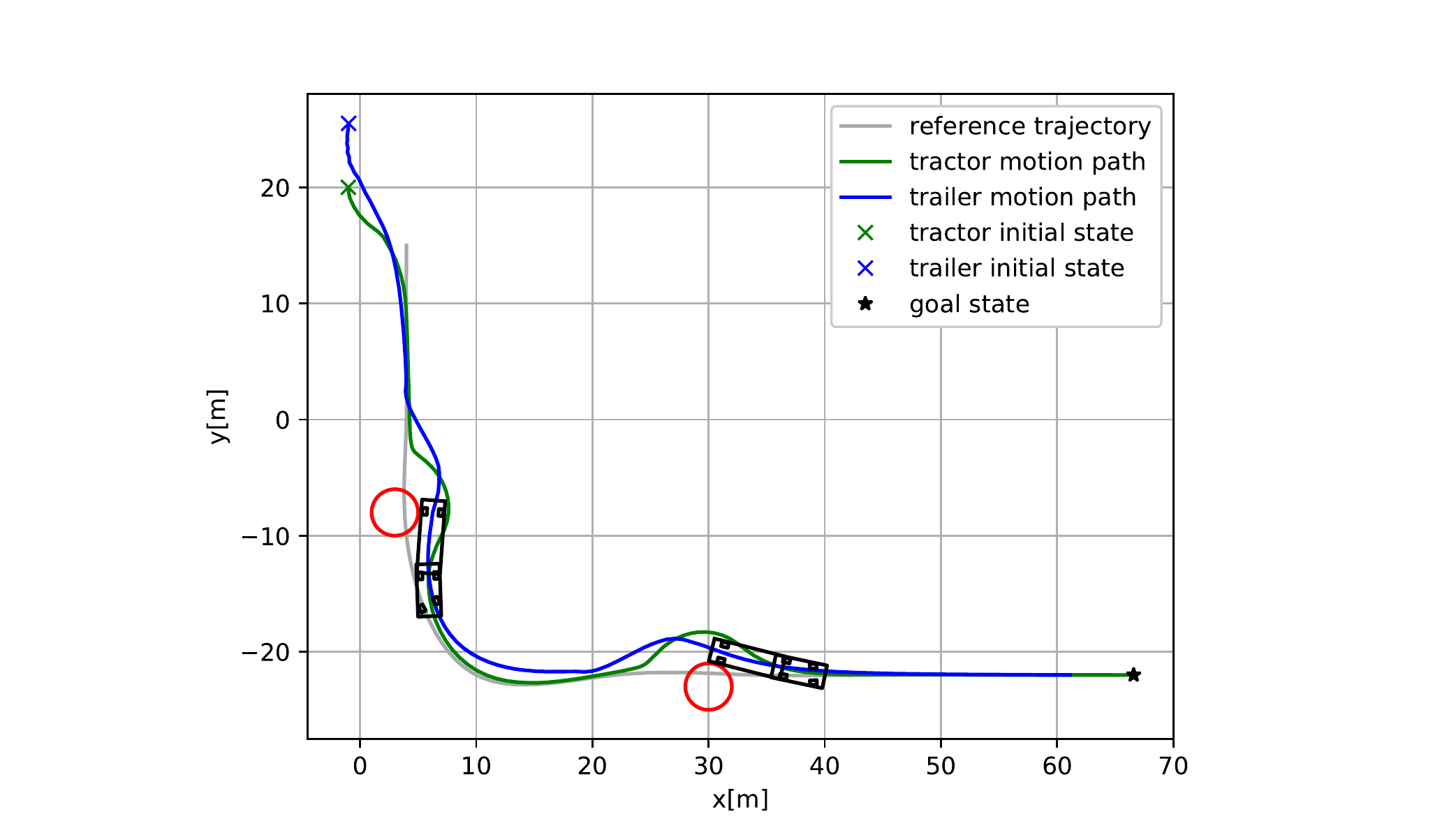}}
\caption{SSTTR with ECBF in a $90^{\circ}$ turning maneuver in the presnece of two obstacles. See https://youtu.be/eQ1NxBjofmU for the simulation video.}
\label{fig2}
\end{figure}

\begin{figure}[htbp]
  \centering%
    \includegraphics[width=.49\linewidth]{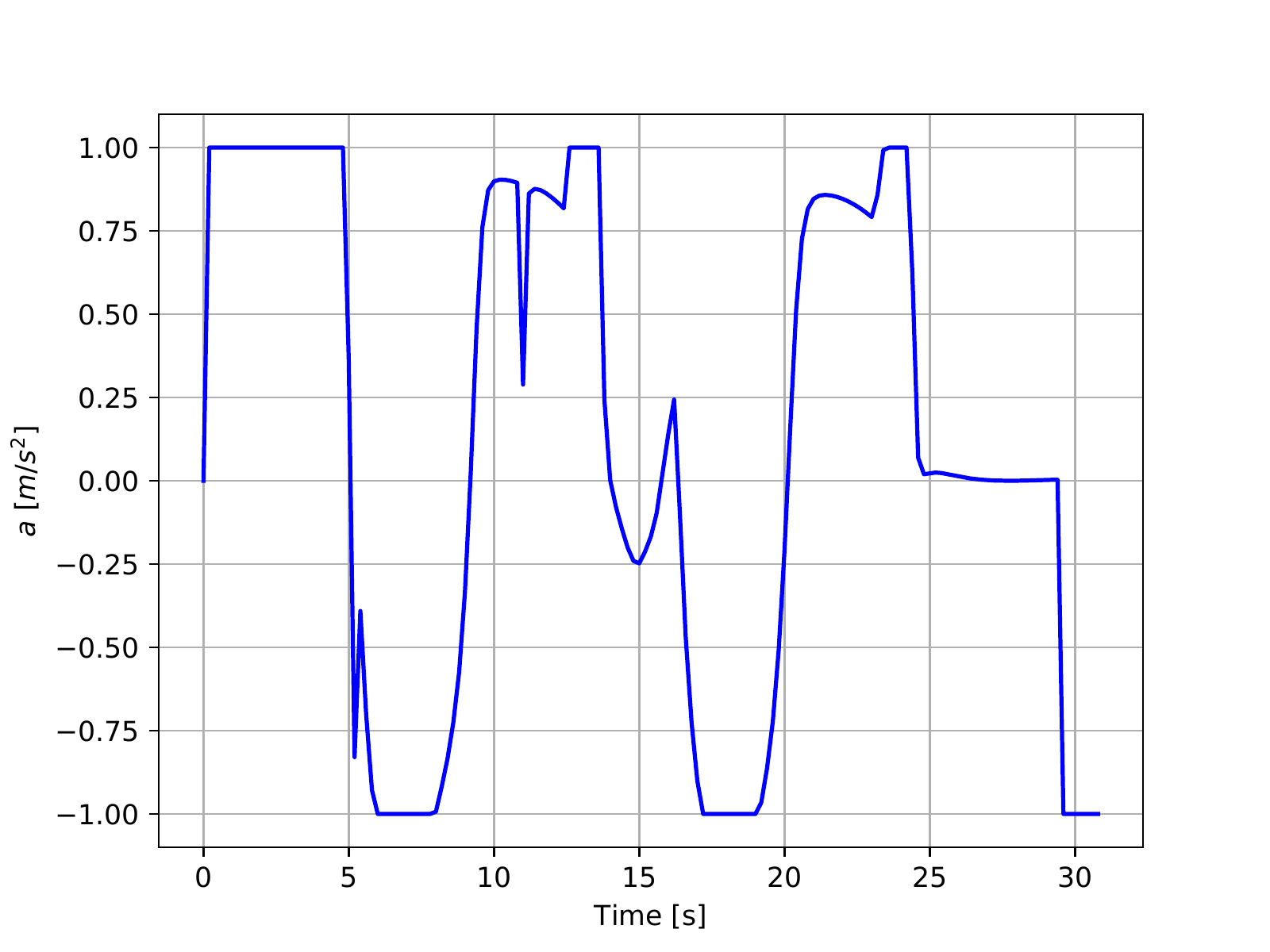}\hfill%
    \includegraphics[width=.49\linewidth]{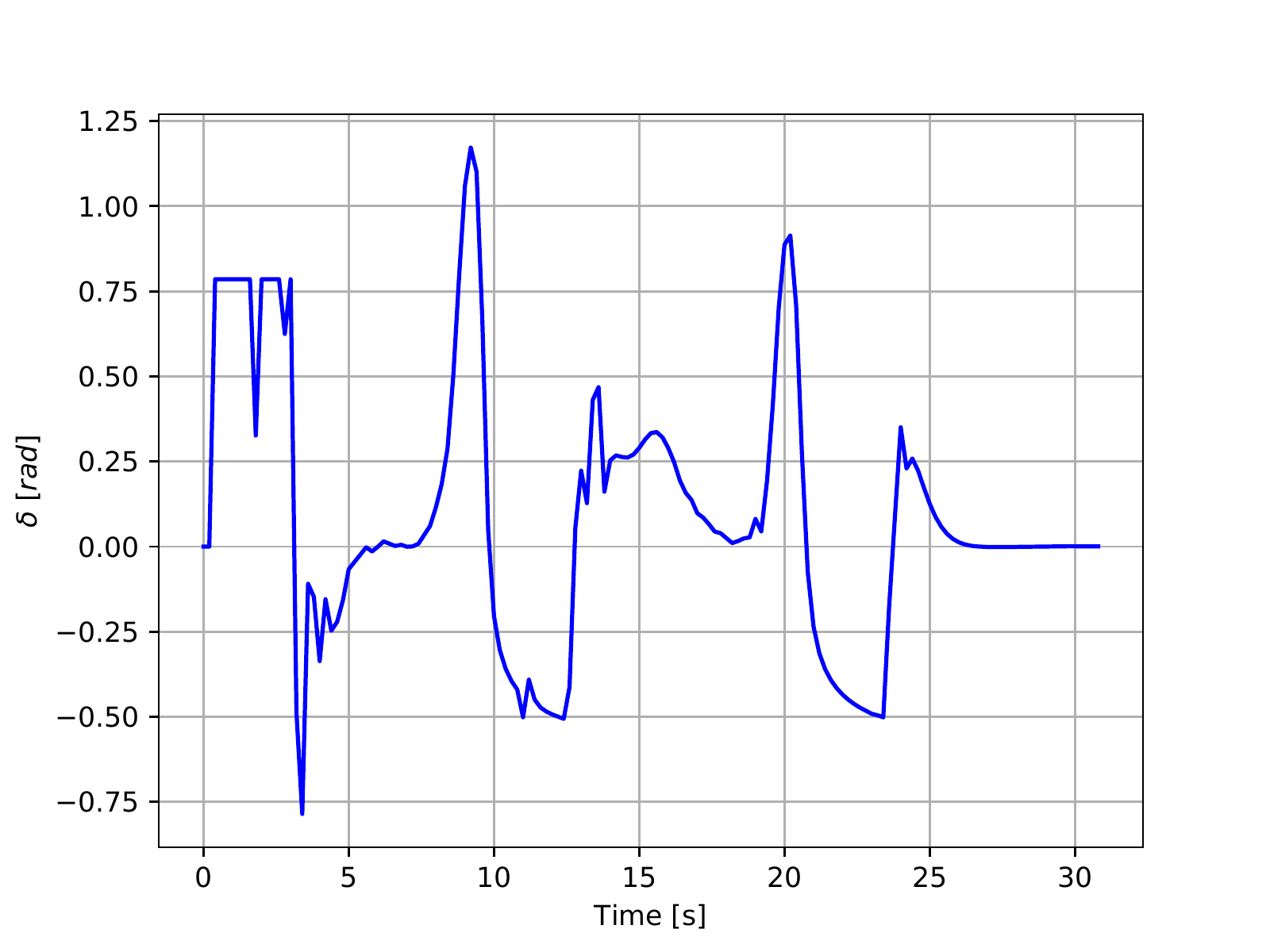}%
    \caption{The SSTTR system inputs. Left: the tractor acceleration. Right: the tractor steering angle.}
    \label{fig:SSTTR_Comparison_inputs}
\end{figure}

\begin{figure}[htbp]
\centerline{\includegraphics[trim=0.2cm 8.0cm 0.2cm 8.0cm, clip, totalheight=0.15\textheight]{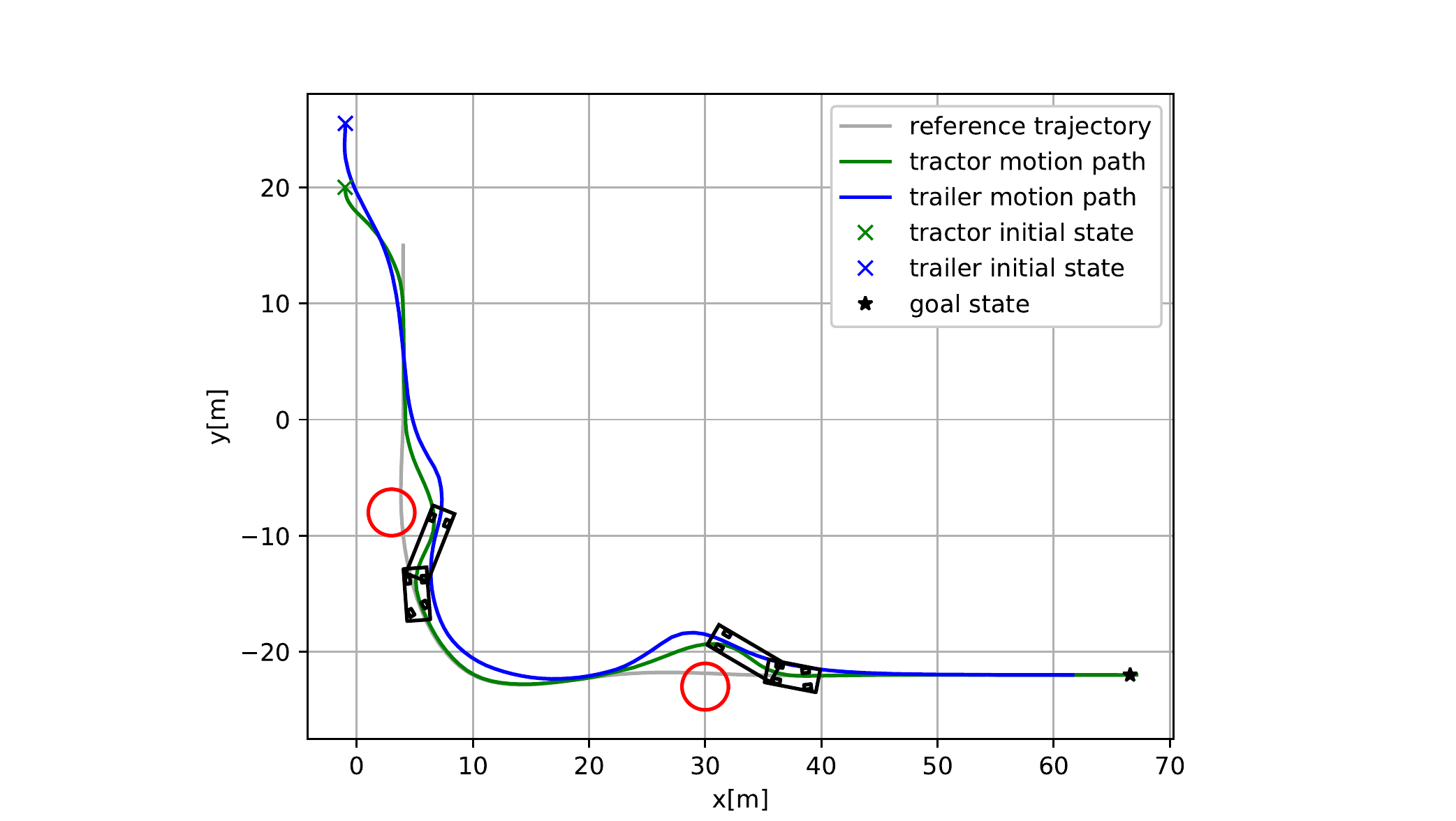}}
\caption{MSTTR with multiple CBFs in a $90^{\circ}$ turning maneuver in the presence of two obstacles. See https://youtu.be/eQ1NxBjofmU for the simulation video.}
\label{fig3}
\end{figure}

\begin{figure}[htbp]
  \centering%
    \includegraphics[width=.33\linewidth]{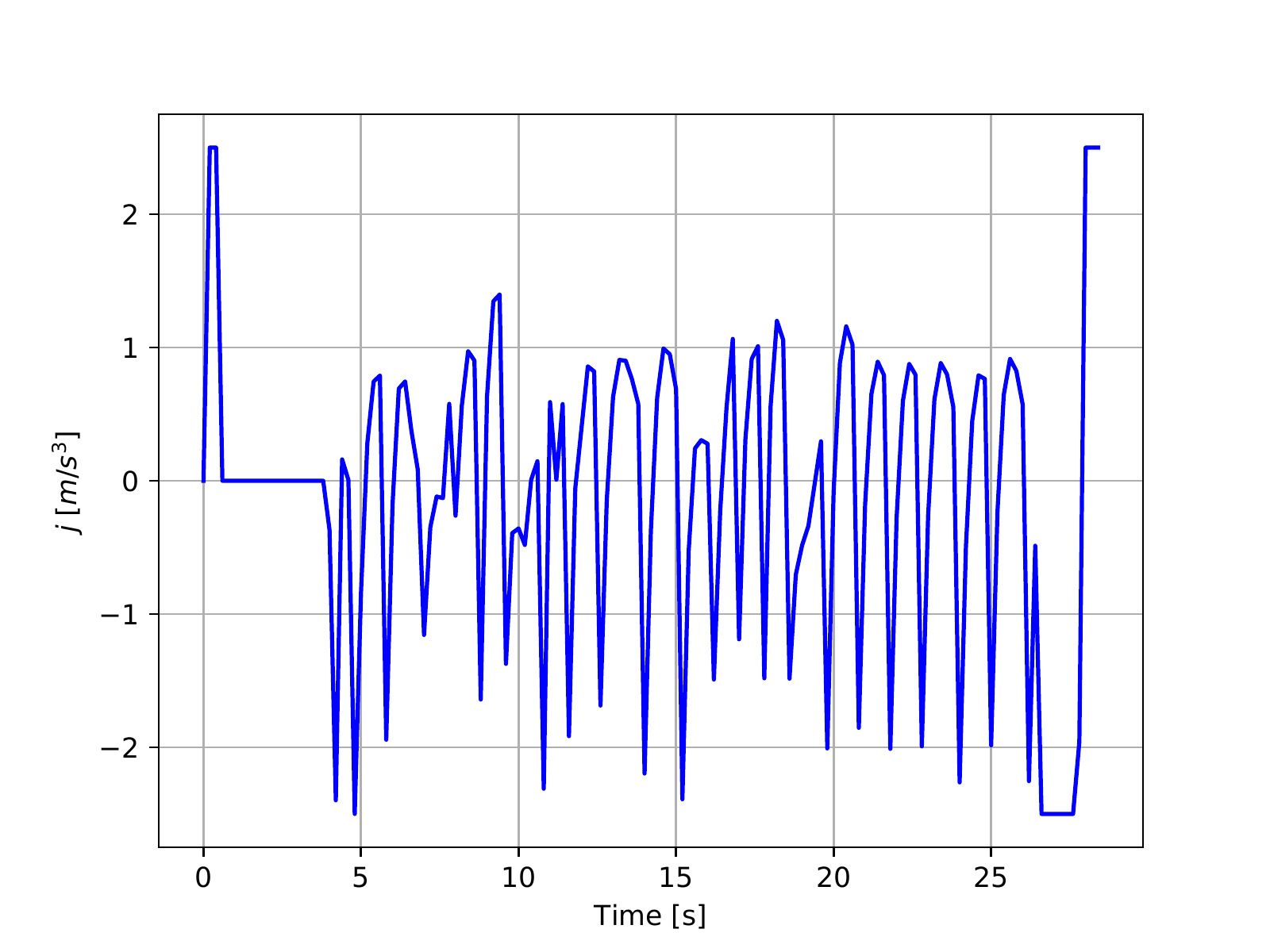}\hfill%
    \includegraphics[width=.33\linewidth]{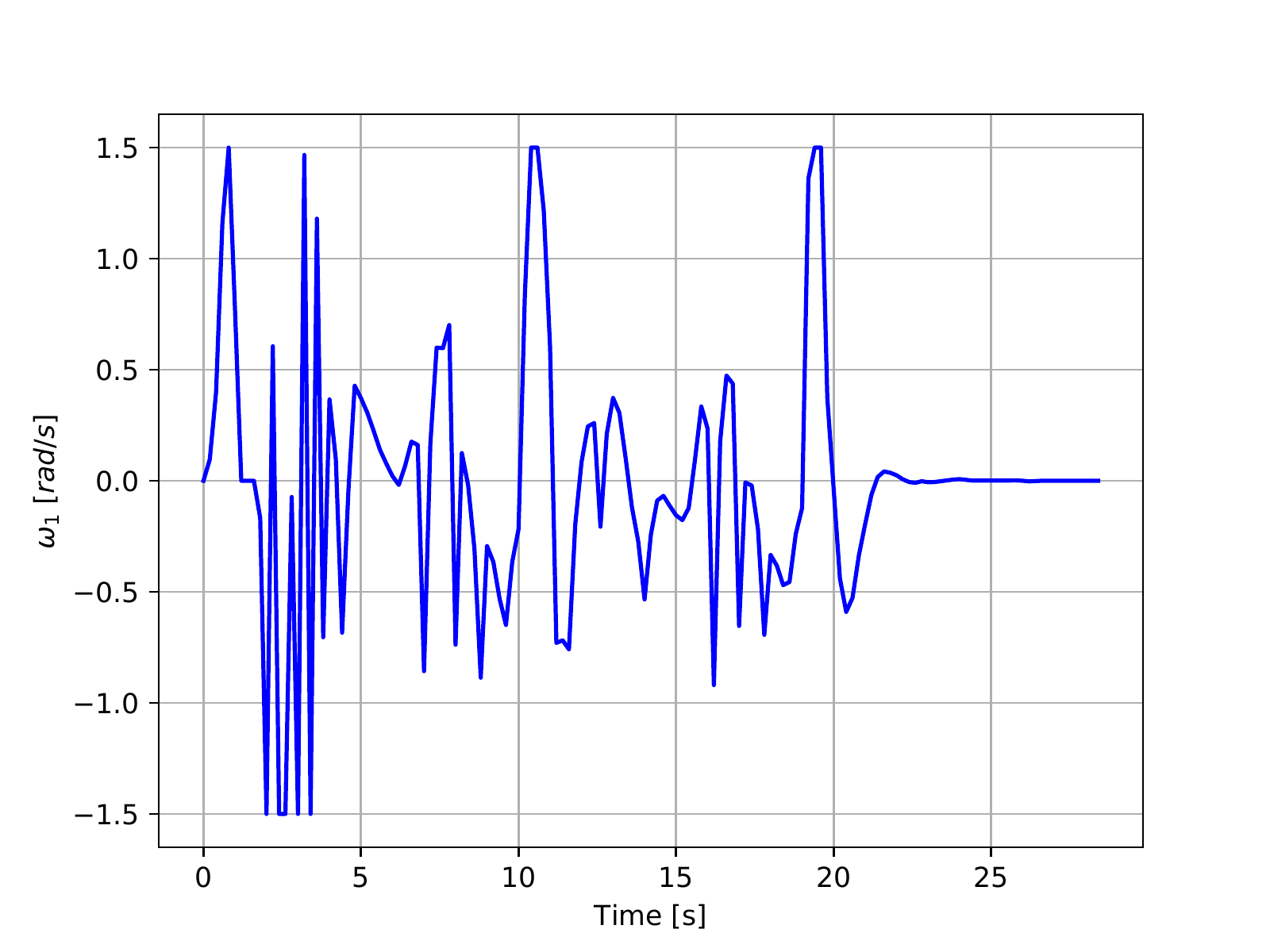}\hfill%
    \includegraphics[width=.33\linewidth]{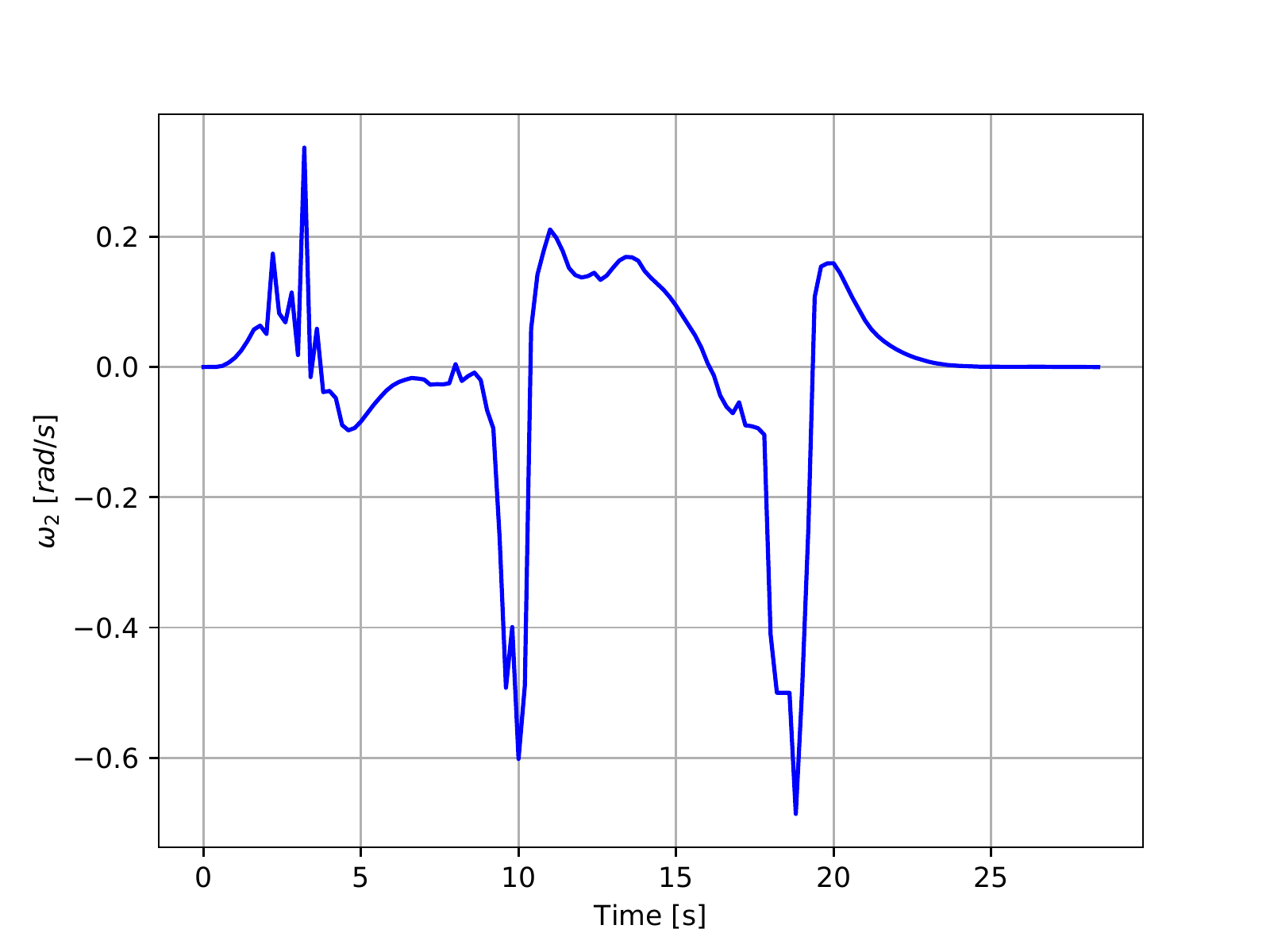}%
    \caption{The MSTTR system inputs. Left: the tractor jerk signal. Middle: the tractor angular velocity. Right: the trailer angular velocity.}
    \label{fig:MSTTR_Comparison_inputs}
\end{figure}

Figure \ref{fig2} and \ref{fig3} illustrate the performance of the ECBF and multiple CBFs, respectively. The transit paths of the tractor and trailer are represented, and the reference trajectory is shown in dark grey. The system started at an arbitrary initial position, where it is not in collision with any obstacle. The results show that the LTV MPC was able to track the reference trajectory in a short amount of time. In Figures \ref{fig2} and \ref{fig3}, the tractor-trailer system is illustrated in two snapshots when it passes each obstacle. As shown in the Figure \ref{fig2}, the SSTTR with ECBF had a collision with obstacles at the corner of the trailer. In contrast, Figure \ref{fig3} shows that the MSTTR with multiple CBFs scheme successfully passed the obstacles without any collision. The reason is that the multiple CBFs approach has a dedicated CBF for each body. It effectively benefits from the MSTTR's additional input to navigate the trailer such that it also passes the obstacle without any collision.
The input signals for the SSTTR and MSTTR systems are illustrated in Figures \ref{fig:SSTTR_Comparison_inputs} and \ref{fig:MSTTR_Comparison_inputs}, respectively. It can be verified that the input constraints are satisfied in both cases using the LTV MPC structure. The additional input in the MSTTR (the angular velocity $\omega_2$) has been represented in Figure \ref{fig:MSTTR_Comparison_inputs}, which shows that the multi-CBFs scheme avoids any collision by changing this input at two time instances when the trailer is passing the obstacles.

\subsection{Obstacle avoidance in a cluttered workspace}
\begin{figure}[htbp]
  \centering
    \includegraphics[width=.49\linewidth]{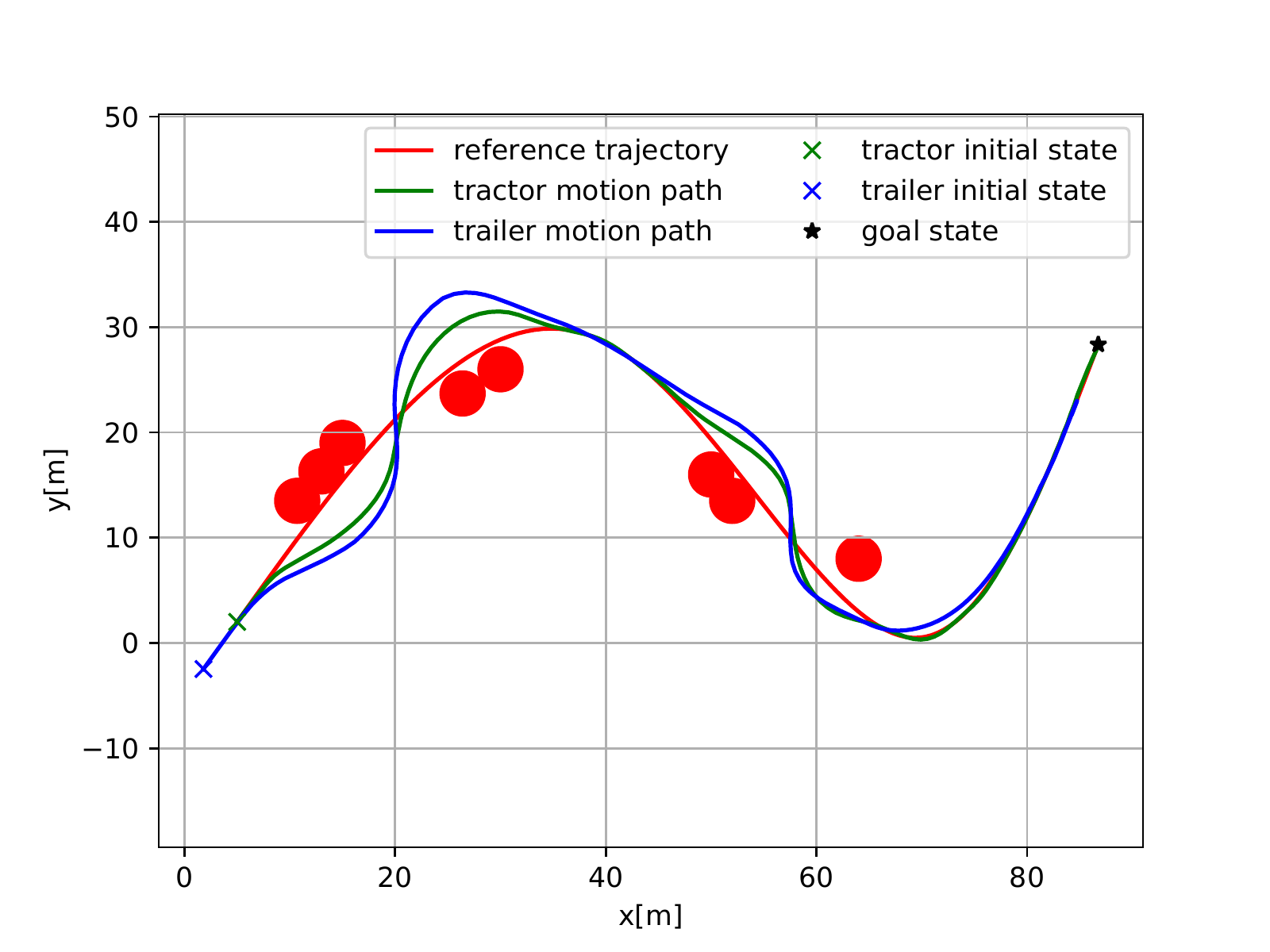}\hfill
    \includegraphics[width=.49\linewidth]{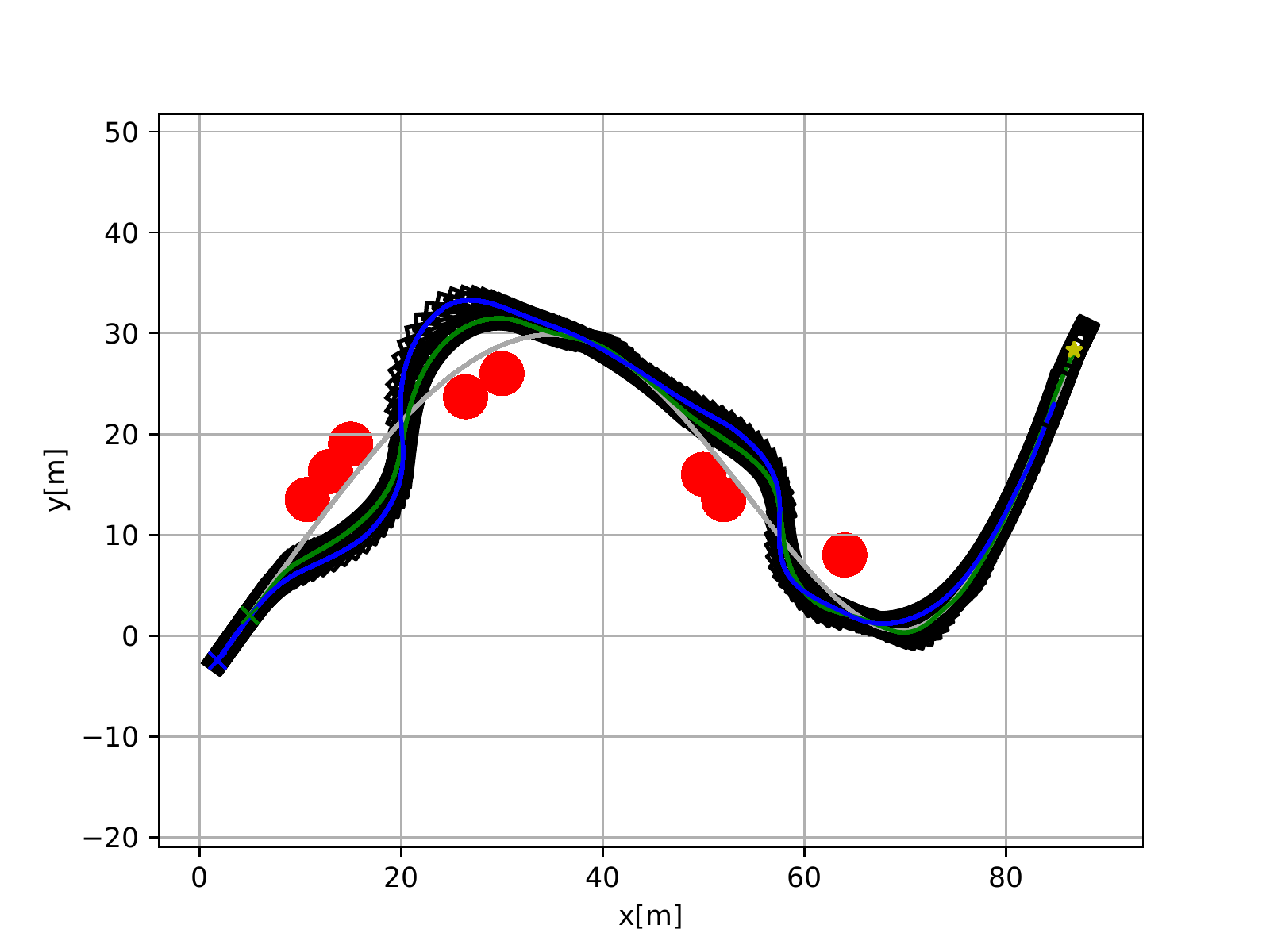}
    \caption{MSTTR with multiple CBFs in a cluttered environment. Left: the trajectories for the tractor and trailer. Right: footprint plot that shows the space covered by the whole system (tractor and trailer) during the simulation. See https://youtu.be/-1xBkKiIoe8 for the simulation video.}
    \label{fig4}
\end{figure}

\begin{figure}
  \centering%
    \includegraphics[width=.33\linewidth]{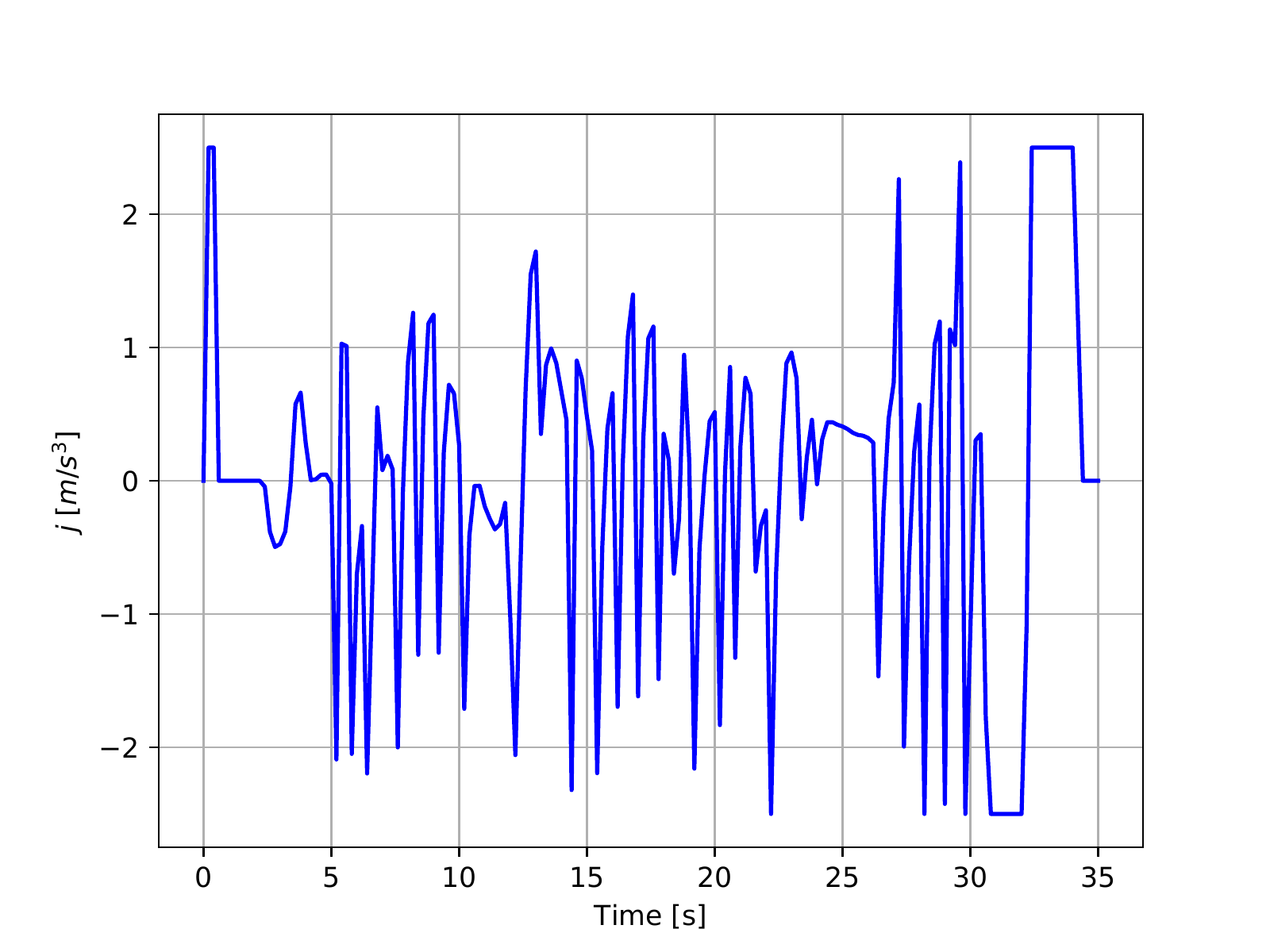}\hfill%
    \includegraphics[width=.33\linewidth]{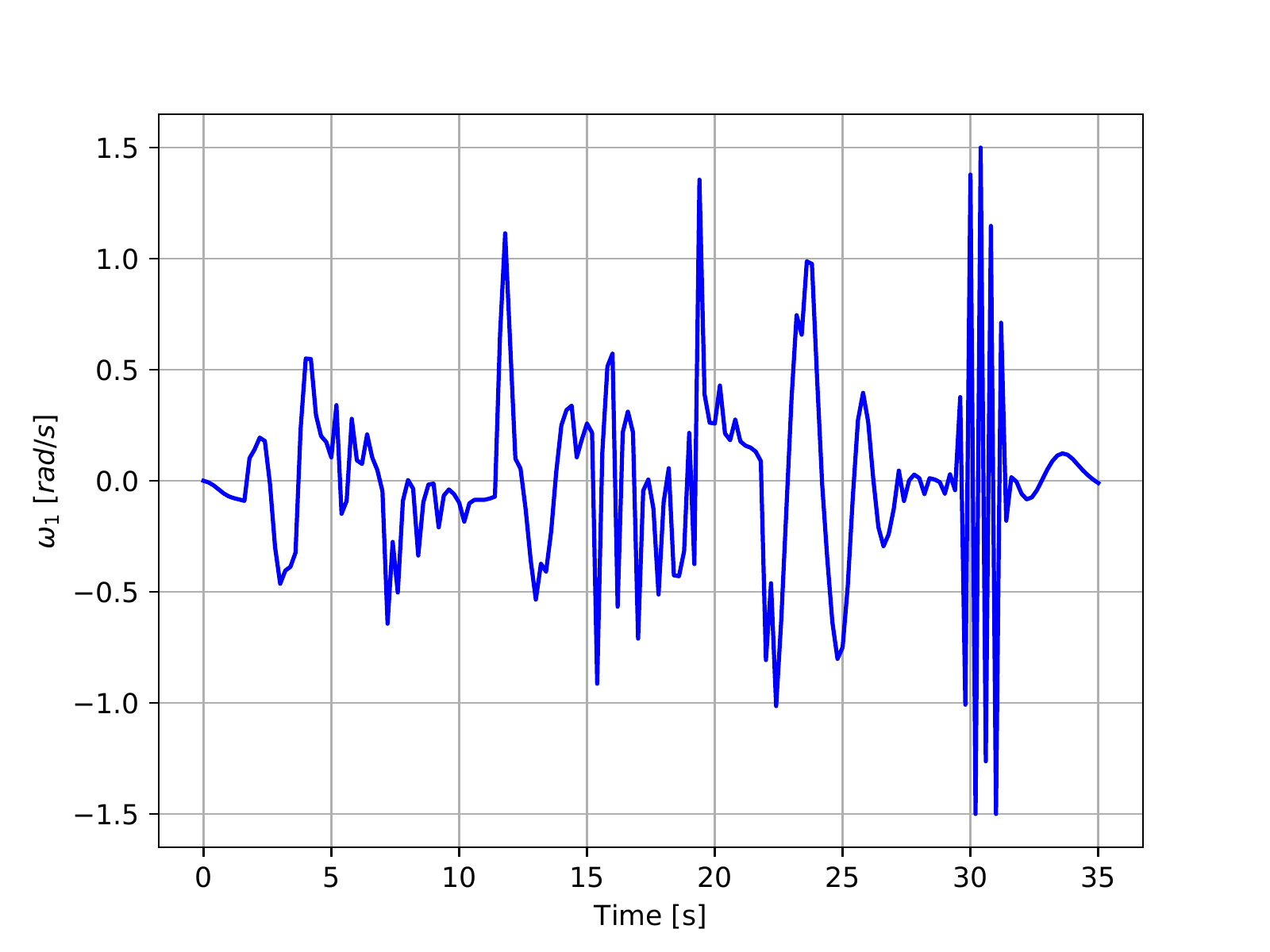}\hfill%
    \includegraphics[width=.33\linewidth]{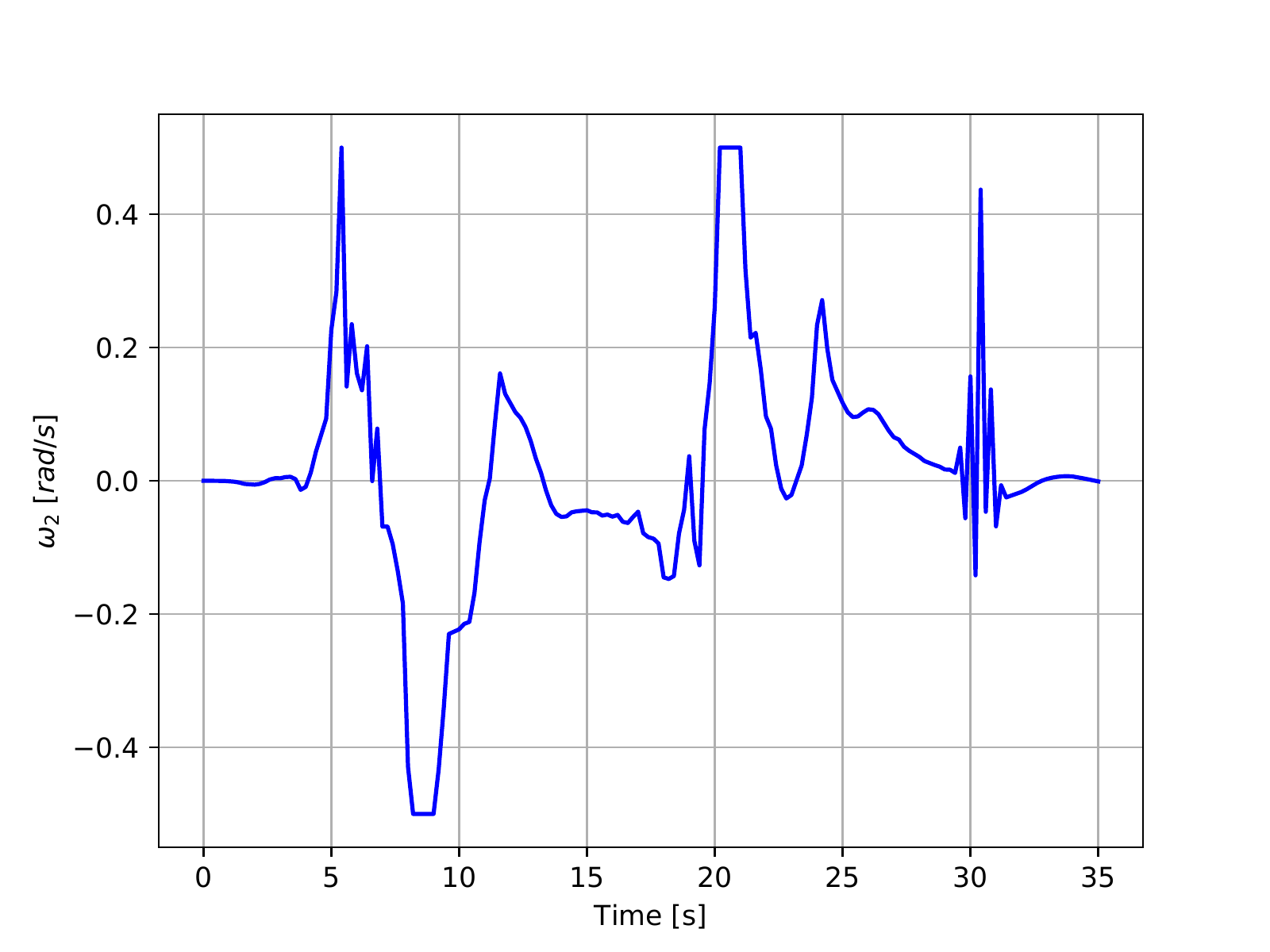}%
    \caption{The MSTTR system inputs in a cluttered environment simulation. Left: the tractor jerk signal. Middle: the tractor angular velocity. Right: the trailer angular velocity.}
    \label{fig:MSTTR_Cluttered_inputs}
\end{figure}

In this section, we investigate the accuracy of the proposed tracking and obstacle avoidance in a complicated scenario. A cluttered workspace is considered where eight obstacles are presented on a reference trajectory. The LTV MPC and multiple CBFs scheme parameters are the same as previously stated. In order to show that the system passed the obstacles without any collision, a footprint plot is provided in Figure \ref{fig4}, which shows the space covered by the whole system (tractor and trailer) during the simulation. The input signals associated with the MSTTR system have also been represented in Figure \ref{fig:MSTTR_Cluttered_inputs}.

The evaluation process in this environment can be divided into two phases. In the first phase, the reactive obstacle avoidance of the proposed method is tested. Particularly, the location of the obstacles left a narrow space for the tractor and trailer to pass without collision. However, the safety module could successfully ensure a collision-free pass while effectively enforcing the safe control input via the QP framework. In time instances that the system passes the obstacles, the input $\omega_2$ changes such that the trailer does not have any collision with the obstacles (see Figure \ref{fig:MSTTR_Cluttered_inputs}). In the second phase, the tracking objective is evaluated. As shown in Figure \ref{fig4}, when the reference is free of obstacles, both the tractor and trailer accurately follow the reference path, which is illustrated in the final part of the simulation.
\section{CONCLUSION}\label{sec6}
Robots are usually exposed to various safety hazards in real-world industrial applications. We must consider these hazards' risk in the control design to achieve a fully autonomous robotic system. In this work, we have introduced a novel multiple CBFs structure, which effectively uses the idea of high-order control barrier functions to ensure the forward invariance of the associated safe sets. A reactive obstacle avoidance technique is designed for MSTTRs based on the proposed multiple CBFs scheme. Simulation results show that the multi-CBFs method ensures collision avoidance for both the tractor and trailer.

Future work will focus on connecting high-level specifications with low-level control design. Checking the feasibility of the QP optimization problem in the presence of multiple CBFs can also be another topic in this area.

\bibliographystyle{IEEEtran}
\bibliography{References}
\section*{Appendix}
\subsection{Tractor and trailer's positional safety constraint}
We need to obtain the following derivatives of $h_{1k}(\boldsymbol{x})$ to assemble (\ref{eq27}). Substituting the derivatives of the position variables $x_1$ and $y_1$ from (\ref{eq23}) up to order three yields
\begin{align}
    \dot h_{1k}(\boldsymbol{x}) ={}& 2v [d_{k}^x c\theta +d_{k}^y s\theta],\nonumber\\
    \ddot h_{1k}(\boldsymbol{x}) ={}& 2v^2 + 2a [d_{k}^x c\theta +d_{k}^y s\theta] -\nonumber\\ 
    &2 \frac{v^2}{l_1} t \delta_1 [d_{k}^x s\theta - d_{k}^y c\theta],\nonumber\\
    \dddot h_{1k}(\boldsymbol{x},\boldsymbol{u}) ={}& 6va - \frac{6va}{l_1} t\delta_1[d_{k}^x s\theta -d_{k}^y c\theta] -2 \frac{v^3}{l_1^2} t^2 \delta_1 \nonumber\\
    &[d_{k}^x c\theta + d_{k}^y s\theta] +2[d_{k}^x c\theta +d_{k}^y s\theta] J- \nonumber\\ 
    &2 \frac{v^2}{l_1}sec^2 \delta_1[d_{k}^x s\theta -d_{k}^y c\theta] \omega_1.\nonumber
\end{align}
Similarly, the required derivatives of $h_{2k}(\boldsymbol{x})$ are obtained by considering (\ref{eq24}) and substituting the derivatives of $x_2$ and $y_2$ up to order two in terms of the tractor's reference point coordinates as follows
\begin{align}
    \dot h_{2k}(\boldsymbol{x}) ={}& 2v[d_{k}^x c\theta +d_{k}^y s \theta] -2 l_2 v c\psi + 2v (t\delta_2 c\psi + s \psi)\nonumber\\ 
    &[d_{k}^x s(\theta - \psi) + d_{k}^y c(\theta - \psi)],\nonumber\\
    \ddot h_{2k}(\boldsymbol{x},\boldsymbol{u}) ={}& 2v^2 -2v^2 (t\delta_2 c\psi+s \psi) s\psi - 2 l_2 a c\psi +\nonumber\\
    & 2 l_2 v s \psi [\frac{v}{l_1} t\delta_1 - \frac{v}{l_2} (t\delta_2 c\psi + s\psi)] +\nonumber\\
    &2a [d_{k}^x c\theta +d_{k}^y s\theta] -2 \frac{v^2}{l_1} t\delta_1 [d_{k}^x s\theta - d_{k}^y c\theta] +\nonumber\\
    & 2a (t\delta_2 c\psi+s \psi) [d_{k}^x s(\theta - \psi) -d_{k}^y c(\theta - \psi)]-\nonumber\\
    & 2v (t\delta_2 s\psi-c\psi) [\frac{v}{l_1} t\delta_1 - \frac{v}{l_2} (t\delta_2 c\psi +s\psi)]\nonumber\\
    &[d_{k}^x s(\theta - \psi) - d_{k}^y c(\theta - \psi)] +\nonumber\\
    & \frac{2v^2}{l_2}(t\delta_2 c\psi+s \psi)^2[d_{k}^x c(\theta - \psi) + d_{k}^y s (\theta - \psi)] +\nonumber\\
    & 2v \sec^2 \delta_2 c\psi [d_{k}^x s(\theta - \psi) - d_{k}^y c(\theta - \psi)] \omega_2,
    \nonumber
 \end{align}
 where $s(\cdot)=\sin(\cdot)$, $c(\cdot)=\cos(\cdot)$, and $t(\cdot) = \tan(\cdot)$. We also denote $(x_1 - x^o_k)$ by $d_{k}^x$ and $(y_1 - y^o_k)$ by $d_{k}^y$ for $k \in \{1,\dots,N_o\}$, respectively.
\subsection{Discrete linear time-varying system}
\begin{align*}
\scriptsize
    A_k= \begin{bmatrix}
1 & 0 & \cos \theta_r T_s & 0 & -v_r \sin \theta_r T_s & 0 & 0 & 0 \\
0 & 1 & \sin \theta_r T_s & 0 & v_r \cos \theta_r T_s & 0 & 0 & 0 \\
0 & 0 & 1 & T_s & 0 & 0 & 0 & 0 \\
0 & 0 & 0 & 1 & 0 
& 0 & 0 & 0 \\
0 & 0 & \frac{\tan \delta_{1r}}{l_1}T_s & 0 & 1 & 0 & a_{57} & 0 \\
0 & 0 & a_{63} & 0 & 0 & a_{66}& a_{67} & a_{68} \\
0 & 0 & 0 & 0 & 0 & 0 & 1 & 0 \\
0 & 0 & 0 & 0 & 0 & 0 & 0 & 1 
\end{bmatrix},
\end{align*}
where $a_{57} = \frac{v_r}{l_1}(1+ \tan^2 \delta_{1r})T_s$, $a_{63} = (\frac{\tan \delta_{1r}}{l_1} - \frac{(\tan \delta_{2r} \cos \psi_r + \sin \psi_r)}{l_2})T_s $, $ a_{66}= 1 - \frac{v_r}{l_2}(-\tan \delta_{2r} \sin \psi_r + \cos \psi_r) T_s $, $a_{67} = \frac{v_r}{l_1}(1+ \tan^2 \delta_{1r})T_s$, and $a_{68}= -\frac{v_r}{l_2}(1+ \tan^2 \delta_{2r})\cos \psi_r T_s$.
\begin{equation*}
\scriptsize
    B_k = 
    \begin{bmatrix}
0 & 0 & 0 \\
0 & 0 & 0 \\
0 & 0 & 0 \\
T_s & 0 & 0 \\
0 & 0 & 0 \\
0 & 0 & 0 \\
0 & T_s & 0 \\
0 & 0 & T_s 
\end{bmatrix},C_k = 
\begin{bmatrix}
v_r \theta_r \sin \theta_r  \\
-v_r \theta_r \cos \theta_r  \\
0 \\
0 \\
c_{51} \\
c_{61} \\
0 \\
0 
\end{bmatrix},
\end{equation*}
where $c_{51} =- \frac{v_r}{l_1}(1+ \tan^2 \delta_{1r}) \delta_{1r} T_s $, and $c_{61} = \delta_{1r}) \delta_{1r} + \frac{v_r}{l_2} \cos \psi_r (1 + \tan^2 \delta_{2r}) \delta_{2r})T_s$.
\end{document}